\documentclass[12pt]{article}%
\usepackage{amssymb}
\usepackage{amsmath}
\usepackage{amsfonts}
\usepackage{graphicx}%
\providecommand{\U}[1]{\protect\rule{.1in}{.1in}}

\begin{document}
\clearpage

\begin{center}
Mimetic Horava Gravity and Surface terms
\end{center}
\vspace{0.5cm}
\begin{center}
O. Malaeb, C. Saghir \\
Physics Department, American University of Beirut, Lebanon
\end{center}
\vspace{3cm}
\begin{abstract}
We consider mimetic Horava gravity, where the scalar field of mimetic gravity was used in the construction of diffeomorphism invariant models reducing to Horava gravity in the synchronous gauge. It will be shown that the surface terms resulting from the variation of the action constructed will cancel out; therefore, there is no need for the addition of Gibbons–Hawking–York boundary term. The resulting surface terms contain higher order space derivatives and no higher order time derivatives.
\end{abstract}

\clearpage


General relativity GR is a classical theory describing gravity through Einstein field equation. It has proven success through several phenomena: the perihelion of mercury, prediction of black holes, gravitational waves... Although of the huge success that GR has achieved, still it is unable to explain some points: dark matter, singularities like the big bang and the black holes... This issue has pushed physicists to modify GR in several ways. Some modified theories of gravity have aimed to quantize GR as a way to explain Big bang singularities or the black holes. Approaches to quantum gravity can be classified into two groups \cite{Claus}.The first group starts from the classical theory of gravity and then applies the quantization steps. The covariant quantum gravity is a good example, where one starts from the path integral of the GR action and then applies the perturbation procedures for the metric around a background one. 
The resulting theory has been a non renormalizable one upon extending it to higher energies. The canonical quantum gravity is another model that belongs to the same group. In this approach, one start constructing the Hamiltonian of the classical GR theory and then turn into the quantum one by quantizing the constraints. Special attention should be taken upon dealing with surface terms \cite{alan}. The second group aims to construct a unified quantum theory of all the forces as in string theory. Restoring renormalizability, by adding higher order derivative terms like $R_{\mu \nu}R^{\mu \nu}$ ends with theories containing ghosts \cite{higherorder}.  The emergence of ghosts was mainly due to the presence of higher order time derivatives. To solve the problem of ghosts, breaking Lorentz invariance at the UV limit was a successful solution.\\ 

The most notable attempt to quantize gravity was the theory of Horava gravity \cite{horava}. Horava chose to break Lorentz invariance by including higher space derivative terms in the Lagrangian while keeping the time derivatives up to second order. Although he got a model that is renormalizable, the low energy limit of the theory was not attained and the negative energy particles reappeared again \cite{horava1}. In attempts to make the model covariant by adding one new field, the property of renormalizability was lost \cite{barv}.  \\

Other modified GR theories aimed to explain dark matter like the mimetic dark matter model proposed by Chamseddine and Mukhanov \cite{mimetic}. Their model is based on the idea of isolation of the scalar degree of freedom of the metric. Not only it was able to predict dark matter but also predicting several cosmological solutions \cite{cosmo} and dark energy \cite{inhomo}, resolving singularities \cite{resol}, \cite{black} and  building a ghost free massive gravity model \cite{mimass}.\\

The Horava renormalizable gravity suffers from a ghost without mimetic. Recently, a mimetic Horava gravity model was proposed by Chamseddine, Mukhanov and Russ. Their model which can regenerate the Horava gravity, in a diffeomorphism invariant way, without the introduction of ghost-like degrees of freedom \cite{mimhorava}.The idea is to construct, within mimetic gravity, all the terms needed in Horava gravity using four-dimensional tensors. These terms will reduce to the wanted form in the synchronous gauge. The scalar field of mimetic gravity $\phi$ was used to write quantities, without introducing new propagating degrees of freedom, invariant under space diffeomorphisms.\\

The mimetic Horava action constructed is given by
\begin{align}
    I= \int \sqrt{-g}( \nabla_{\mu} \nabla_{\nu} \phi \nabla^{\mu} \nabla^{\nu} \phi - (\Box \phi)^2 + \tilde{R})d^4 x
    \label{action}
\end{align}
where $\tilde{R} = 2R_{\mu \nu} \partial_{\mu} \phi \partial_{\nu} \phi - R -(\Box \phi)^2 + \nabla_{\mu} \nabla {\nu} \phi \nabla^{\mu} \nabla^{\nu} \phi$.\paragraph{}
  
In this letter, we aim to prove that the surface terms coming from GR will be canceled by the surface terms resulting from the added terms in the mimetic Horava action. The main goal behind canceling the surface term is to perform the Hamiltonian analysis which is the base for canonical quantization of GR.  \paragraph{}

Considering first the surface terms emerging in GR, upon variation of the Hilbert-Einstein action with respect to the metric, a surface integral over the boundary $\partial M$ is obtained. The surface integral resulting is non-vanishing. Therefore, in the Hamiltonian quantization of gravity, including boundary terms in the action is a must. This has been known since years ago \cite{Gibbon}. The inclusion of boundary terms results in obtaining a variational principle that is well defined and yields the Einstein field equations. This manipulation was needed because a second order derivative hides in the Ricci scalar R, which is not allowed in the path integral formalism. To get an action that depends only on the first derivatives of the metric, these second derivatives terms are removed by integration by parts. Then the resulting action is quadratic in first derivatives of the metric. The surface terms resulting are canceled by amending the action to \cite{horowitz}, \cite{euclidean}
\begin{align}
    I= - \frac{1}{16 \pi} \int_M d^4 x \sqrt{g} R  - \frac{1}{8 \pi} \int_{ \partial M} d^3 x \sqrt{h} K
\end{align}
where $\partial M$ is the boundary of M, $h_{ab}$ is the induced metric on $\partial M$, and K is the trace of the second fundamental form on $\partial M$. In canonical formulation, some ignore the presence of the surface term. As a result, certain boundary terms must be added to the constraint to give a well defined equations of motion \cite{Regge}. \paragraph{}
The remarkable point that we will show here is that starting from the mimetic Horava action, the signs and the coefficients of the surface terms will be just right to cancel out. Therefore, there is no need for the Gibbons Hawking boundary term to be added. The mimetic Horava action (\ref{action}) can be rewritten as 
\begin{align}
    I= \int \sqrt{-g}( -R - 2\nabla_{\mu}(\Box \phi \nabla^{\mu}\phi) +2 \nabla_{\sigma}( \nabla_{\mu} \nabla^{\sigma} \phi \nabla^{\mu} \phi))d^4 x
\end{align}\\

We start by the Einstein Hilbert action
\begin{align}
I_H= \int\sqrt{-g} R d^4 x
\end{align}

\begin{align}
    \delta I_H= \int_{\gamma} G_{\alpha \beta} \delta g^{\alpha \beta} \sqrt{-g} d^4x  - \oint_{\partial \gamma} \epsilon h^{\alpha \beta} \delta g_{\alpha \beta , \mu} n^{\mu} |h|^{\frac{1}{2}} d^3 y 
    \label{surface}
\end{align}
Where 
\begin{align}
    n_{\mu}n^{\mu}= \epsilon \\
    g^{\mu \nu}= \epsilon n^{\mu}n^{\nu}+h^{\mu \nu} 
\end{align}

Let
\begin{align}
    I_1 = \int \sqrt{-g}(-2\nabla_{\mu}(\Box \phi \nabla ^{\mu} \phi)d^4 x)
\end{align}
and
\begin{align}
    I_2= \int \sqrt{-g} (2 \nabla_{\sigma}( \nabla_{\mu} \nabla^{\sigma} \phi \nabla^{\mu} \phi)d^4 x)
\end{align}
The variation of $I_1$ gives
\begin{align}
    \delta I_1 = I_1 '(\delta \phi) + 2 \oint_{\partial \gamma} \nabla_{\mu} \phi \frac{1}{2} g^{\rho \sigma} g^{\alpha \lambda} (2 \delta g_{\rho \lambda, \sigma}- \delta g_{\rho \sigma, \lambda}) \partial_{\alpha} \phi d\Sigma^{\mu} 
\end{align}
Where 
\begin{align}
    I_1'(\delta \phi)= 2 \oint_{\partial \gamma} \nabla_{\mu} \phi \Box \delta \phi d \Sigma^{\mu} + 2 \oint_{\partial \gamma} \partial_{\mu} \delta \phi \Box \phi d\Sigma^{\mu}.
\end{align}
The variation of $I_2$ term gives 
\begin{align}
    \delta I_2 = I'_2(\delta \phi) -2 \oint_{\partial \gamma} g^{\rho \tau}( \delta g_{\mu \tau, \sigma}+ \delta g_{\sigma \tau, \mu} - \delta g_{\mu \sigma, \tau}) \nabla_{\rho} \phi \nabla^{\mu} \phi d\Sigma^{\sigma}
\end{align}

Where
\begin{align}
    I_2'(\delta \phi)= 2 \oint_{\partial \gamma} \nabla_{\mu} \nabla_{\sigma} \delta \phi \nabla^{\mu} \phi d\Sigma^{\sigma} + 2\oint_{\partial \gamma} \nabla_{\mu} \nabla_{\sigma} \phi \nabla^{\mu} \delta \phi d\Sigma^{\sigma}
\end{align}\\
The variation in phi terms, $I_1'(\delta \phi)$ and $I_2'(\delta \phi)$, could be integrated out (surface of a surface) and will vanish upon setting $\delta \phi = 0$ on the boundaries. \\
The question is under what condition (if any) the surface terms, over $\delta g_{\mu \nu}$, of $\delta I_H$, $ \delta I_1$ and $\delta I_2$ will cancel out upon addition. Using 
\begin{align}
    d\Sigma^{\mu}= n^{\mu} \epsilon |h|^{\frac{1}{2}} d^3 y
\end{align}
along with the completeness relation of the metric 

\begin{align}
    g^{\alpha \beta}= \epsilon n^{\alpha} n^{\beta} + h^{\alpha \beta}
\end{align}
it turns out that the surface terms will cancel out upon the choice of 
\begin{align}
    n_{\mu}=\partial_{\mu} \phi.
\end{align}
    
As a summary, quantizing gravity helps to explain several unsolved problems by GR. Most quantum gravity theory depends on the Hamiltonian analysis of GR. The presence of a surface term in the action of GR, generates problems in the equations of motion. Mimetic Horava gravity is a new model which regenerate the Horava gravity without any ghost. Making use of the vector $n_{\mu}= \partial_{\mu}\phi$, any Lagrangian can be projected to have higher space derivative terms and no higher time derivative ones. The variation of the resulting higher space derivative surface terms is capable of canceling the surface term resulting from the variation of Einstein Hilbert action. This proves that the Gibbon's Hawking boundary term are not the only surface terms that can be used to cancel the surface terms resulting from the variation of the Einstein Hilbert action. Getting rid of the surface terms makes the 3+1 splitting of the theory much easier. 

\clearpage

\textbf{{\large {Acknowledgments}}}

We would like to thank Professor Ali Chamseddine for suggesting the problem and for his helpful discussions on the subject. We would like to thank also the American University of Beirut (Faculty of Science) and the National Council for Scientific Research of Lebanon (CNRS-L) for granting a doctoral fellowship to (Chireen Saghir) and for their support.

\end{document}